\documentclass[aps,pra,showpacs,twocolumn]{revtex4}
\usepackage{mathrsfs}
\usepackage{amssymb}
\usepackage{amsmath}
\usepackage{bm}
\usepackage{graphics}
\usepackage{natbib}
\usepackage{color}

\begin{document}
\title{${\cal PT}$ symmetry of the Su-Schrieffer-Heeger model with imaginary boundary potentials and  next-nearest-neighboring coupling}

\author{Xue-Si Li}

\author{Ze-Zhong Li}

\author{Lian-Lian Zhang}

\author{Wei-Jiang Gong}\email{gwj@mail.neu.edu.cn}

\affiliation{College of Sciences, Northeastern University, Shenyang 110819, China}
\date{\today}

\begin{abstract}
By introducing the next-nearest-neighboring (NNN) intersite coupling, we investigate the eigenenergies of the $\cal PT$-symmetric non-Hermitian Su-Schrieffer-Heeger (SSH) model with two conjugated imaginary potentials at the end sites. It is found that with the strengthening of NNN coupling, the particle-hole symmetry is destroyed. As a result, the bonding band is first narrowed and then undergoes the top-bottom reversal followed by the its width's increase, whereas the antibonding band is widened monotonously. In this process, the topological state extends into the topologically-trivial region, and its energy departs from the energy zero point, accompanied by the emergence of one new topological state in this region. All these results give rise to the complication of the topological properties and the manner of $\cal PT$-symmetry breaking. It can be concluded that the NNN coupling takes important effects to the change of the topological properties of the non-Hermitian SSH system.
\end{abstract}
\pacs{11.30.Er, 03.65.Vf, 73.21.Cd}
\maketitle

\bigskip

\section{Introduction}
Over the past decades, systems with non-Hermitian Hamiltonians have become an important research concern in the quantum-physics-related field. In particular, the Hamiltonians with parity-time ($\cal PT$) symmetry are more noticeable. The main reason is due to that below the phase transition point of symmetry breaking, the systems have an opportunity to possess purely real eigenenergy spectra\cite{Bender}. Many researches have shown that $\cal PT$ symmetry presents very fruitful contents, which bring about the fundamental physics and abundant applications in many fields. Moreover, the advances and developments in experiment allow $\cal PT$-symmetric non-Hermitian systems to be optically achieved by incorporating gain and loss\cite{Nature}. And the signatures and potential applications of them have been demonstrated in various aspects, such as optical waveguides\cite{Natphys,yhuang,snghosh,yfu,szhang}, perfect cavity absorber lasers\cite{slonghi,ydchong,HJS}, single-mode lasing action in $\cal PT$-symmetric microcavity arrangements\cite{mamiri,hhodaei}, microwave cavities and resonators\cite{hhodaei2,bpeng}, particularly in metamaterials with extraordinary properties\cite{yaure,dwang,gptsi,mkang,yxu,ysun}. In addition, some systems with $\cal PT$-symmetric Hamiltonians have been investigated. For instance, the optoelectronic oscillators have been designed for $\cal PT$ symmetry in hybrid optoelectronic systems\cite{YanzhongLiu,JieJunZhang}.
\par
It is well known that in optical experiments, topological systems can be constructed and studied in details, even if they are difficult to realize in the field of solid-state physics\cite{top1,top2}. Photons in coupled waveguides and optical lattices are manipulated in the manner similar to the electrons in solids, providing intriguing opportunities for novel optical devices\cite{top3,top4,top5}. Accordingly, this promotes the establishment of a new field, namely, topological photonics. And some important work have been reported. The topologically protected unidirectional interface state has been experimentally demonstrated in coupled waveguide ring resonators\cite{top6}. The photonic topological insulator have been realized by embedding an anisotropic photonic crystal into a metallic plate waveguide\cite{top6a}. Researchers also report a new type of phononic crystals with topologically nontrivial band gaps for both longitudinal and transverse polarizations, resulting in protected one-way elastic edge waves\cite{top7}. And other groups have demonstrated that synthesizing artificial gauge fields for ultracold atoms in optical lattices enables the construction of a two-dimensional topological system\cite{top8,top9}.
\par
The progress of both $\cal PT$-symmetric optics and topological photonics directly induces the development of the topic of $\cal PT$-symmetric topological systems\cite{toppt1,toppt2,toppt3,toppt3a}. One reason is because of its underlying new physics, and the other is for that theoretical anticipations can be verified in a relatively short period\cite{YoungsunChoi,Ti01,Ti02,Ti03,before1}. According to the previous works, topologically protected $\cal PT$-symmetric interface states have been demonstrated in coupled resonators\cite{toppt4,toppt5}, and $\cal PT$-symmetric non-Hermitian Aubry-Andr\'{e} systems\cite{toppt6} and Kitaev models\cite{toppt7,toppt7a} have been theoretically investigated. Relevant conclusions show that universal non-Hermiticity can alter topological regions\cite{toppt8,toppt9,toppt9a}, but topological properties are robust against local non-Hermiticity. Also, $\cal PT$-symmetric discrete-time quantum walk has been realized, with which edge states between regions with different topological numbers and their robustness to perturbations and static disorder have been observed\cite{LXiao}. Latest report began to focus on the two-dimensional $\cal PT$-symmetric system, which shows that the non-Hermitian two-dimensional topological phase transition coincides with the emergence of mid-gap edge states by means of photonic waveguide lattices with judiciously designed refractive index landscape and alternating loss\cite{Kremer}.
\par
\par
As the simplest topological model, the Su-Schrieffer-Heeger (SSH) chain certainly attracts extensive attentions, by considering the $\cal PT$ symmetry different aspects. Zhu $et$ $al$. have studied the $\cal PT$-symmetric non-Hermitian SSH model with two conjugated imaginary potentials at two end sites. They find that the non-Hermitian terms can lead to different effects on the properties of the eigenvalue spectra in topologically
trivial and nontrivial phases. And in the topologically trivial phase, the system undergoes an abrupt transition from the
unbroken $\cal PT$-symmetry region to the spontaneously broken $\cal PT$-symmetry region at a certain potential-magnitude $\gamma_c$\cite{Zhu1}. After this work, many groups dedicate themselves to investigating the properties of $\cal PT$-symmetric non-Hermitian SSH model, including the spontaneous $\cal PT$-symmetry breaking, topological invariants, topological phase, harmonic oscillation at the exceptional point (EP), and topological end states\cite{SSH1,SSH2,SSH3,SSH4,SSH5,SSH6}. In addition, the effects of defects on the topological states in the non-Hermitian SSH system have been already considered.
And more complicated systems have begun to receive attentions, e.g., the
SSH and Kitaev models, non-Hermitian trimerized lattices, and tetramerizd systems\cite{Kitaev,Triple,PRL}.
\par
Following the research progress, we would like to investigate the eigenenergies of the $\cal PT$-symmetric non-Hermitian SSH model with two conjugated imaginary potentials at the end sites, by introducing finite next-nearest-neighboring (NNN) intersite coupling. Our purpose is to clarify the special influences of particle-hole symmetry breaking induced by the presence of NNN coupling. The calculation results show that the band structure of this system undergoes substantial changes. As for the topological state, it does not only show the energy shift from the zero point but also extends into the \emph{original} topologically-trivial region. Moreover, new topological state arises in such a region. All these results exactly lead to substantial changes in the topological properties and the $\cal PT$-symmetry breaking of the non-Hermitian SSH system, in the presence of NNN intersite coupling.
\par

\begin{figure}
\centering \scalebox{0.41}{\includegraphics{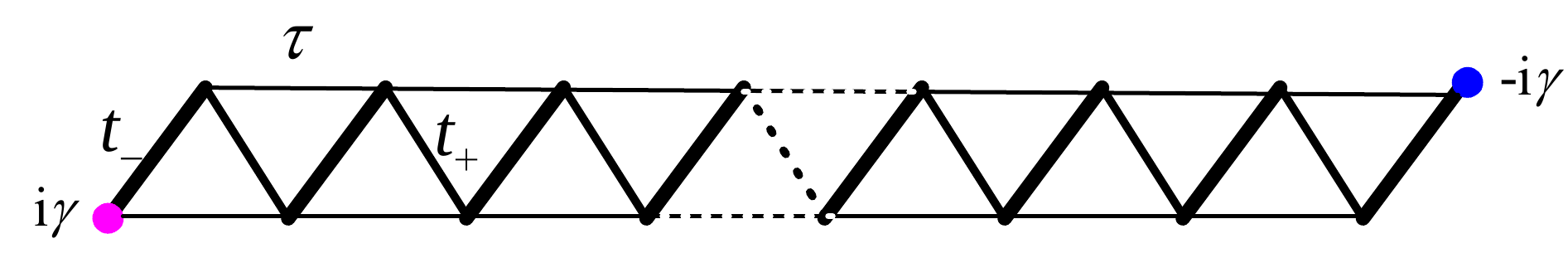}} \caption{
Schematic of the $\cal PT$-symmetric non-Hermitian SSH model with NNN coupling. Two conjugated imaginary on-site potentials are introduced at its end sites to mimic the non-Hermitian terms. The nearest intersite couplings are labeled as $t_+$ and $t_-$, and the NNN intersite coupling is taken to be $\tau$.  \label{structure}}
\end{figure}
\section{Theoretical model}
The quasi-1D SSH model that we consider is illustrated in Fig.1, which
is described as a tight-binding chain with alternatingly modulated nearest-neighboring intersite couplings. Besides, the NNN intersite coupling term is taken into account to present its special role in modifying the topological properties of this system. With respect to the non-Hermitian terms, they are achieved by introducing two additional conjugated imaginary on-site potentials at end sites of the chain. The
Hamiltonian of this structure can be written as
\begin{equation}
H=H_{0}+U_{pt},  \label{1}
\end{equation}
where $H_0$ is the SSH model with NNN intersite coupling, given by
\begin{eqnarray}
H_{0}&=&\sum_{j=1}^{N}t_-c^\dag_{2j-1} c_{2j}+\sum_{j=1}^{N-1}t_+c^\dag_{2j} c_{2j+1}+\mathrm{h.c.}\notag\\
&+&\sum_{j=1}^{2N-2}\tau c^\dag_{j} c_{j+2}+\mathrm{h.c.}.\label{2}
\end{eqnarray}
The site number of the SSH chain has been assumed to be $2N$. $c^\dag_j$ ($c_j$) is
the creation (annihilation) operator on the $j$-th site. The nearest neighboring intersite hoppings are given
alternatively by $t_-$ and $t_+$ with $t\mp=t(1\mp\delta \cos\theta)$ to reflect the difference between the two nearest-neighboring intersite hoppings. The parameter $\delta$ is the strength of dimerization, which is greater than zero. And $\theta$
is an introduced tuning parameter, which can vary in one period, i.e., from $-\pi$ to
$\pi$. For convenience, $\delta$ is defined as $|\delta|<1$ and
$t=1$ is set as the unit of energy.
\par
It is known that the
conventional SSH model is the simplest 1D two-band model which enables to exhibit topologically nontrivial properties. To be concrete, the SSH model has the topologically nontrivial phase in the region of $-{\pi\over2} <\theta< {\pi\over2}$ characterized by the presence of zero-mode end states under the open boundary condition, whereas no end states exist in the regions
of $-{3\pi\over2}<\theta<-{\pi\over2}$ (or ${\pi\over2}<\theta<{3\pi\over2}$) corresponding to the topologically
trivial phase. For the SSH model under the periodical
boundary condition, these two phases are distinguished
by the Berry phase, which takes $0$ in the trivial phase and $\pi$ in the nontrivial phase. The zero-mode end states in the nontrivial phase are topologically protected by both the
inversion symmetry and particle-hole symmetry\cite{TI}.
\par
In Hamiltonian $H$, the $U_{pt}$ term describes
two additional conjugated imaginary on-site potentials acting
at the two end sites. It takes the form as
\begin{equation}
U_{pt}=i\gamma c^\dag_1 c_1-i\gamma c^\dag_{2N} c_{2N},
\end{equation}
in which energy gain occurs at the first site and energy loss takes place at the $2N$-th site. Besides, $\gamma$ is the identical strength of the imaginary potentials. In the context, we consider it to be greater than zero.

\par
Since we aim at the study about non-Hermitian SSH model, it is necessary to demonstrate the relevant symmetry properties that can be referred in this work. $\cal P$ and $\cal T$ are the space-reversal (or parity) operator and the time-reversal
operator. Their effects are manifested as $p\to -p$ and $x\to -x$;
$p\to-p$, $x\to x$, and $i\to -i$, respectively. For a $\cal PT$-symmetric Hamiltonian, it should obey the relationship that $[{\cal PT}, H]=0$.
With respect to our considered system with discrete lattice, the effects of $\cal P$ and $\cal T$ are
${\cal P}c_j{\cal P}= c_{2N+1-j}$, and ${\cal T}i{\cal T}=-i$, respectively. It is easy to prove that our non-Hermitian SSH model is $\cal PT$ symmetric, i.e., $({\cal P}{\cal T}) H({\cal T} {\cal P})=H$, since
\begin{eqnarray}
&&{\cal P}H_0{\cal P}=H_0, {\cal T}H_0{\cal T}=H_0,\notag\\
&&{\cal P}U_{pt}{\cal P}=-U_{pt}, {\cal T}U_{pt}{\cal T}=-U_{pt}.
\end{eqnarray}
This exactly means that Hamiltonian $H$ possesses neither $\cal P$ nor $\cal T$ symmetry separately, but it is
invariant under the combined operation of $\cal P$ and $\cal T$.

\par
According to the previous works, the Hamiltonian $H$ can be classified to be either unbroken
$\cal PT$ symmetry or broken $\cal PT$ symmetry, which can be differentiated by observing the symmetry of its eigenfunctions\cite{Bander11,Bander12}. With the help of the Schr\"{o}dinger equation $H|\psi\rangle=E|\psi\rangle$, the eigenfunction $|\psi\rangle$ can be solved. If all the eigenfunctions
have $\cal PT$ symmetry, i.e.,
\begin{equation}
{\cal PT} |\psi\rangle=|\psi\rangle,
\end{equation}
the SSH system will show the unbroken $\cal PT$ symmetry and all
the corresponding eigenvalues are real. Nevertheless, if not all the
eigenfunctions obey the above eigenvalue equation, the system will be of the broken $\cal PT$
symmetry. In such a case, and the complex eigenvalues of the SSH model begin to come into play.

\begin{figure}[htb]
\centering \scalebox{0.27}{\includegraphics{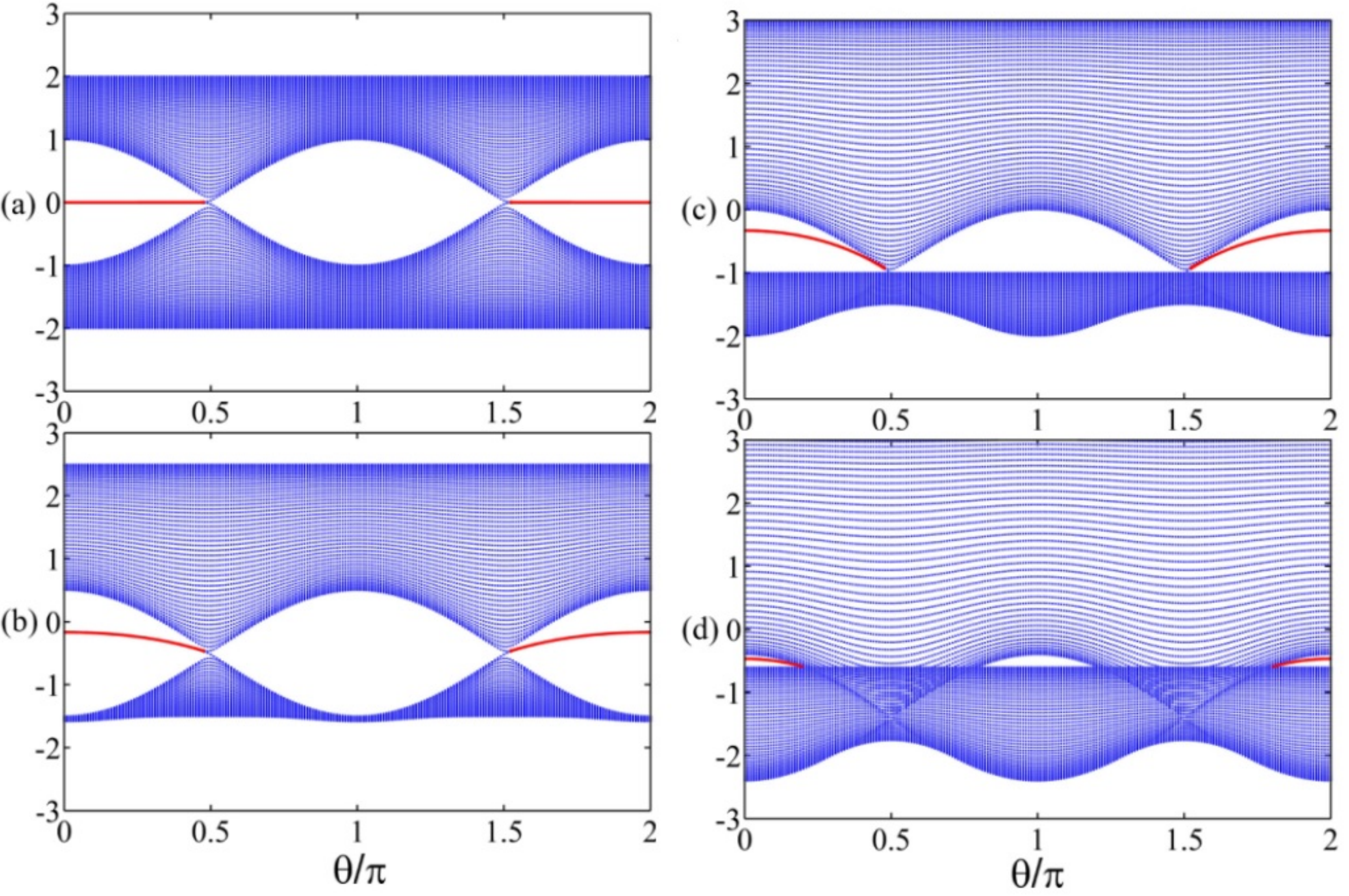}} \caption{
Eigenenergy spectra for the SSH model influenced by the presence of NNN intersite coupling, under the open boundary condition. Relevant parameters are taken to be $\delta=0.5$ and $N=50$. (a) $\tau=0$, (b) $\tau=0.2$, (c) $\tau=0.4$, (d) $\tau=0.7$. }
\end{figure}

\section{Numerical results and discussions \label{result2}}
In this section, we proceed to discuss the characteristics of the eigenenergy spectra of the $\cal PT$-symmetric non-Hermitian SSH model on the basis of the theory that presented in Sec II, with the presence of NNN intersite coupling. Our purpose is to clarify the influence of the NNN intersite coupling on the presence of the topological state and the related $\cal PT$-symmetry breaking. In this work, we take the parameter values as $t=1.0$, $\delta=0.5$, and $N=50$ to perform the numerical calculation.
\par
To begin with, we would like to present the eigenenergy spectra of the Hermitian SSH model with the NNN intersite coupling $\tau$, by ignoring the $\cal PT$-symmetric imaginary potentials. Under such a Hermitian condition, the role of the NNN coupling in modifying the eigenenergies can be clearly seen. The numerical results are shown in Fig.2(a)-(d), where $\tau=0$, $0.2$, $0.4$, and $0.7$, respectively.
In Fig.2(a), It can be observed that in the conventional SSH model, the bonding and antibonding bands encounter at the critical points, i.e., $\theta=\frac{1}{2}\pi$ and $\theta=\frac{3}{2}\pi$. In the regions of $0<\theta<{\pi\over2}$ and  ${3\pi\over2}<\theta<2\pi$, the topologically nontrivial phase comes into being, accompanied by the appearance of zero-energy mode. These results are completely consistent with those in Ref.\cite{Zhu1}.
Once the small NNN coupling is taken into account, the two bands become asymmetric about each other, and the bonding (antibonding) band is narrower (wider), as shown in Fig.2(b). Meanwhile, the energy of the topological state varies to be nonzero and depends on the change of $\theta$. When $\theta$ tunes to the position of $\theta=0$ $(2\pi)$, the energy value of the topological state gets close to the energy zero point. This result turns to be more clear with the increase of $\tau$ to $0.4$ [see Fig.2(c)], followed by the reversed structure of the bonding band. Next in Fig.2(d), for a large $\tau$, e.g., $\tau=0.7$, the two bands overlap with each other, and then the topological state is destroyed accordingly. These results suggest that the NNN coupling takes its destructive effect to the topological state in the SSH model. The underlying reason should be attributed to the breaking of particle-hole symmetry induced by the NNN intersite coupling.
\par
\begin{figure}[htb]
\centering \scalebox{0.088}{\includegraphics{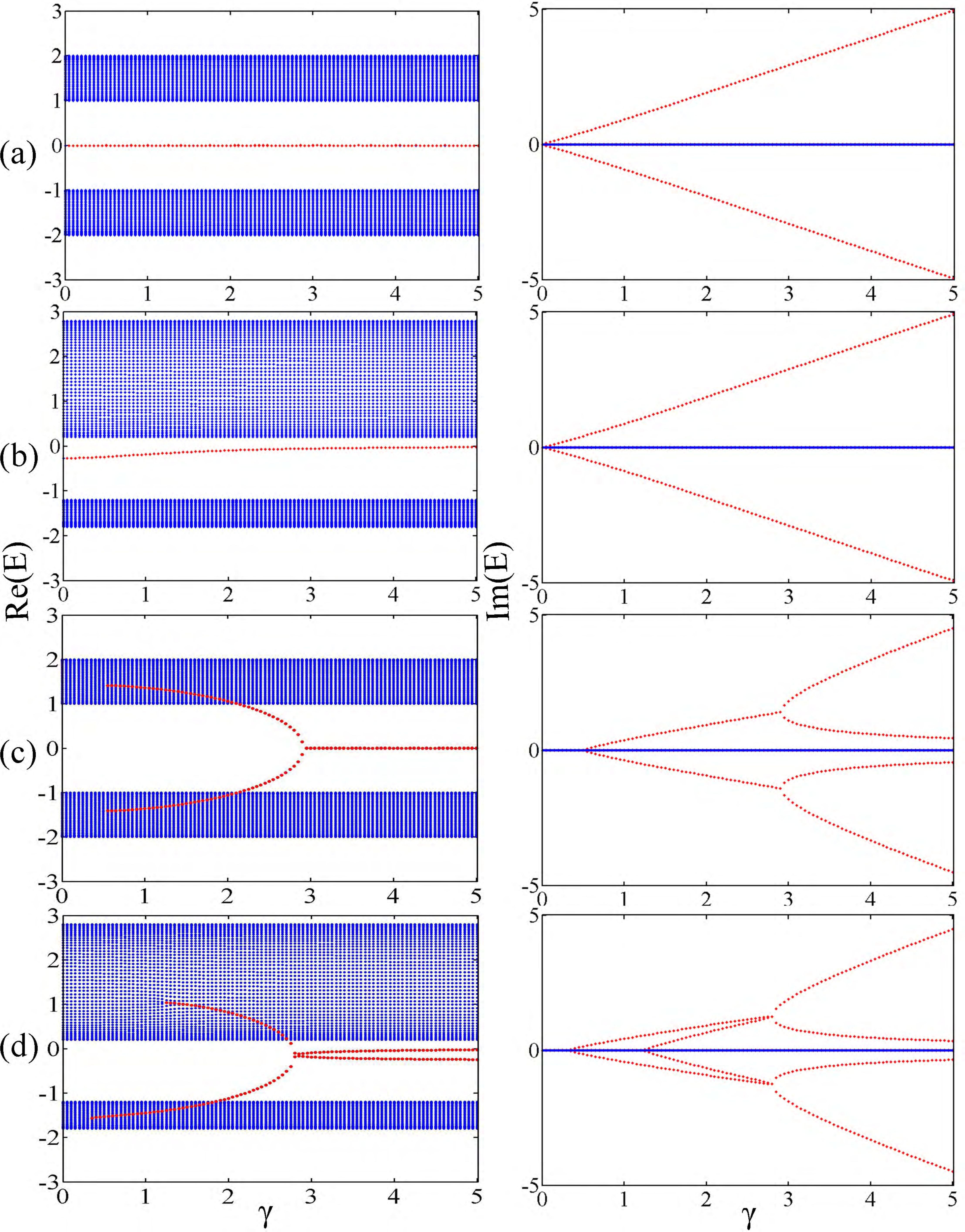}} \caption{
Eigenenergies of the $\cal PT$-symmetric non-Hermitian SSH model as a function of $\gamma$ with $\delta=0.5$ and $N=50$ for finite $\theta$ and $\tau$. (a) $\theta=0$, $\tau=0$; (b) $\theta=0$, $\tau=0.4$; (c) $\theta=\pi$, $\tau=0$; (d) $\theta=\pi$, $\tau=0.4$. Left column describes the real part with the right being the corresponding imaginary part.}
\end{figure}
We next introduce the $\cal PT$-symmetric imaginary potentials to the end sites of the SSH model with finite NNN coupling, to pay attention to the changes of the eigenenergy spectra following the increase of the magnitude of the $\cal PT$-symmetric imaginary potentials (i.e., $\gamma$). The corresponding results are shown in Fig.3. For comparison, the cases of $\tau=0$ are also presented. We find in Fig.3(a) that in the case of $\theta=0$ and $\tau=0$, the real part of the eigenenergy keeps invariant, with the increase of $\gamma$. As for the topological state, the real part of its energy is always equal to zero in this process. However the imaginary part appears in one pair of values throughout the whole range of $\gamma$, which are proportional to the magnitude of imaginary potentials. Next in the presence of NNN coupling with $\tau=0.4$, Fig.3(b) shows that the two bands become asymmetric, and the negative real part of the topological-state energy gets close to zero with the further increment of $\gamma$. In contrast, the imaginary part is the same as the case of $\tau=0$, thus it is independent of the influences of NNN coupling. These phenomena indicate that in the case of $\theta=0$, the topological property and the manner of $\cal PT$-symmetry breaking tends to be weakly dependent on the NNN coupling.
\par
The situations for $\theta=\pi$ are shown in Fig.3(c)-(d), where $\tau=0.0$ and $0.4$. We see that in the case of $\tau=0$, two topological states appear in the bonding and antibonding bands simultaneously, symmetric about the energy zero point, when $\gamma$ arrives at $0.5$. And going on enhancing the imaginary potentials causes these two topological states to degenerate with the EP near the position of $\gamma\approx3.0$. Hereafter, the real part of the topological-state energy keeps equal to zero. The changes of the imaginary part are well consistent with the real part. As the $\cal PT$-symmetric potentials are enlarged to $\gamma\ge0.5$, the imaginary part arises in two pairs of values which are identical with each other, whereas it experiences the second transition in the case of $\gamma\approx3.0$ because the two pairs of imaginary part vary in different manners. They can be viewed as the other kinds of $\cal PT$-symmetry breaking, in comparison with Fig.3(a)-(b). For convenience, the types of $\cal PT$-symmetry breaking should be defined. The case in Fig.3(a)-(b) is defined as type-I, characterised by one pair of imaginary part of energy. And those in Fig.3(c) can be labeled as type-II and type-III, since two pairs of imaginary part come into being with their different $\gamma$ dependence. Alternatively if $\tau=0.4$, Fig.3(d) shows that due to the breaking of particle-hole symmetry, one single topological state arises in the bonding band when $\gamma\approx 0.3$ (the first EP), with one pair of imaginary energies corresponding to it. Thus the type-I $\cal PT$-symmetry breaking comes into being. With the increase of $\gamma$ to $1.3$ (the second EP), the other topological state emerges in the antibonding band, and the $\cal PT$-symmetry breaking enters the type-II stage, where two pairs of imaginary energies exhibit different magnitudes. Next, the further increase of $\gamma$ gives rise to the appearance of the third EP near $\gamma\approx2.7$. However after this EP, the real energies of the two topological states do not exhibit degeneracy, with a little difference between them.
\par
\begin{figure}[htb]
\centering \scalebox{0.050}{\includegraphics{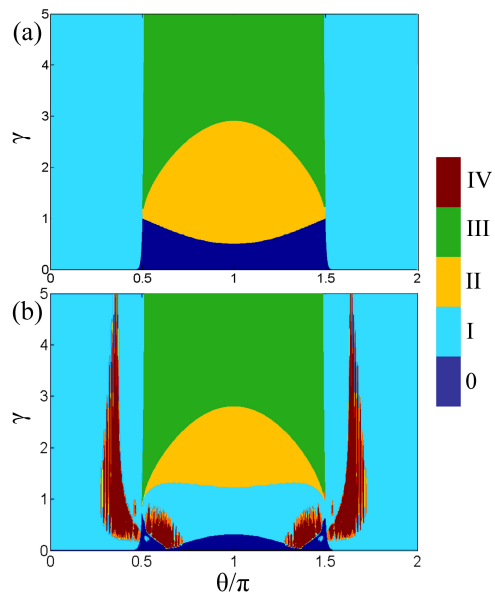}} \caption{Phase diagram that describes the $\cal PT$-symmetry breaking of the SSH model. The parameters are taken to be $\delta=0.5$ and $N=50$. Respective phases are labeled as different colors, which indicate the types of the $\cal PT$-symmetry. (a) $\tau=0$; (b) $\tau=0.4$.}
\label{Fig4}
\end{figure}
\par
Inspired by the result in Fig.3, we would like to plot the phase diagram that describes the $\cal PT$-symmetry breaking for the whole ranges of both $\gamma$ and $\theta$, as shown in Fig.4. In this figure, the I, II, III, and IV phases indicate the types of $\cal PT$-symmetry breaking, and the 0 phase means the absence of this symmetry breaking. Note, also, that the inter-phase boundary exactly corresponds to the EP. In Fig.4(a) it can be readily found that in the cases of $0<\theta<\frac{1}{2}\pi$ and $\frac{3}{2}\pi<\theta<2\pi$, i.e., the \emph{original} topologically nontrivial region, the system is always manifested as the I phase, corresponding to the type-I $\cal PT$-symmetry breaking, which cannot be modulated by the imaginary potentials. While in the region of $\frac{1}{2}\pi<\theta<\frac{3}{2}\pi$ (the so-called \emph{original} topologically-trivial region), the phase transition is tightly determined by the $\cal PT$-symmetric potentials, and three phases are formed with the occurrence of two types of $\cal PT$-symmetry breaking, following the increase of $\gamma$. The EPs obey the relationships that $\gamma\approx1+{1\over2}\cos\theta$ and $\gamma\approx1-2\cos\theta$, respectively. Next, when the NNN intersite coupling is taken into account, one can find the intricate changes of the phase diagram, as shown in Fig.4(b). To be concrete, four phases exist in the region of $\frac{1}{2}\pi<\theta<\frac{3}{2}\pi$, which are related to three types of $\cal PT$-symmetry breaking. We notice that compared with Fig.4(a), the NNN coupling narrows the 0-phase region and divides the II-phase region into the I- and II-phase parts. Thus, with the increase of $\gamma$, the property of the first EP is changed seriously, and new EP appears as well. In addition to the above results, the IV phase emerges in the 0-phase and I-phase regions. This can be attributed to the fact that in the $\cal PT$-symmetric SSH system, the NNN coupling causes the eigenenergies of the bulk states to be complex, and then the type-IV $\cal PT$-symmetry breaking takes places. We then begin to know the role of NNN intersite coupling in driving the topological properties of the SSH model as well as the $\cal PT$-symmetry breaking, due to its-induced particle-hole symmetry.
\par
\begin{figure}[htb]
\centering \scalebox{0.088}{\includegraphics{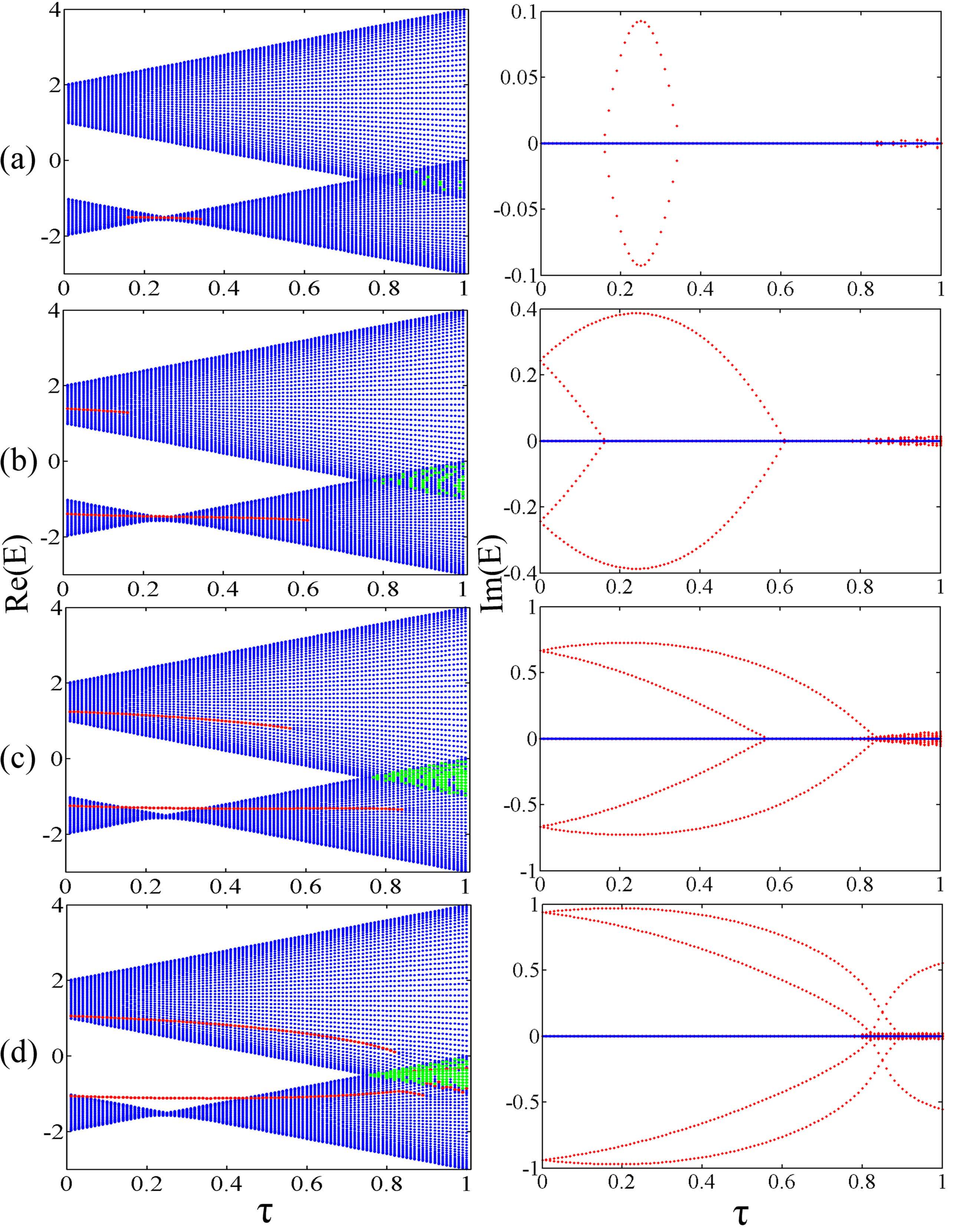}} \caption{
Real and imaginary parts of the eigenenergy spectra as a function of $\tau$ with parameters $ \delta=0.5$, $\theta=\pi$ and $N=50$ for several values of $\gamma$.
(a) $\gamma=0.2$; (b) $\gamma=0.8$; (c) $\gamma=1.5$; and (d) $\gamma=2.0$.}
\label{Fig5}
\end{figure}
In view of the leading results in Fig.4, we would like to increase the NNN coupling to investigate the details and present the corresponding results in Fig.~\ref{Fig5} by taking $\theta=\pi$. As for the $\cal PT$-symmetric potentials, we choose $\gamma=0.2$, $0.8$, $1.5$, and $2.0$, respectively. Let us first focus on the real part of the eigenenergy. One can find that with the increase of $\tau$, the variation manners of the bonding and antibonding bands are still the same as those in Fig.2 where $\gamma=0$. As for the topological states, their presence are tightly relevant to the magnitude of the $\cal PT$-symmetric imaginary potentials. In the presence of $\gamma$, one topological state first arises at the narrow-band limit of the bonding band, which mixes with the bulk state of this band [see Fig.5(a)]. When the larger $\gamma$ is considered, the topological state is allowed to exist in the relatively wide range of $\tau$, meanwhile, the other topological state emerges in the antibonding band, as shown in Fig.5(b)-(d). This result becomes more apparent following the increase of $\gamma$. Therefore, the NNN intersite coupling governs the appearance manner of the topological states, but the strong $\cal PT$-symmetric potentials are key conditions to maintain the topological states. In addition, note that if the NNN coupling causes the band-overlap phenomenon to take place, the eigenenergies of the bulk states exhibit their nonzero imaginary part as well (labeled in the green color), leading to the occurrence of type-IV $\cal PT$-symmetry breaking.

\begin{figure}[htb]
\centering \scalebox{0.16}{\includegraphics{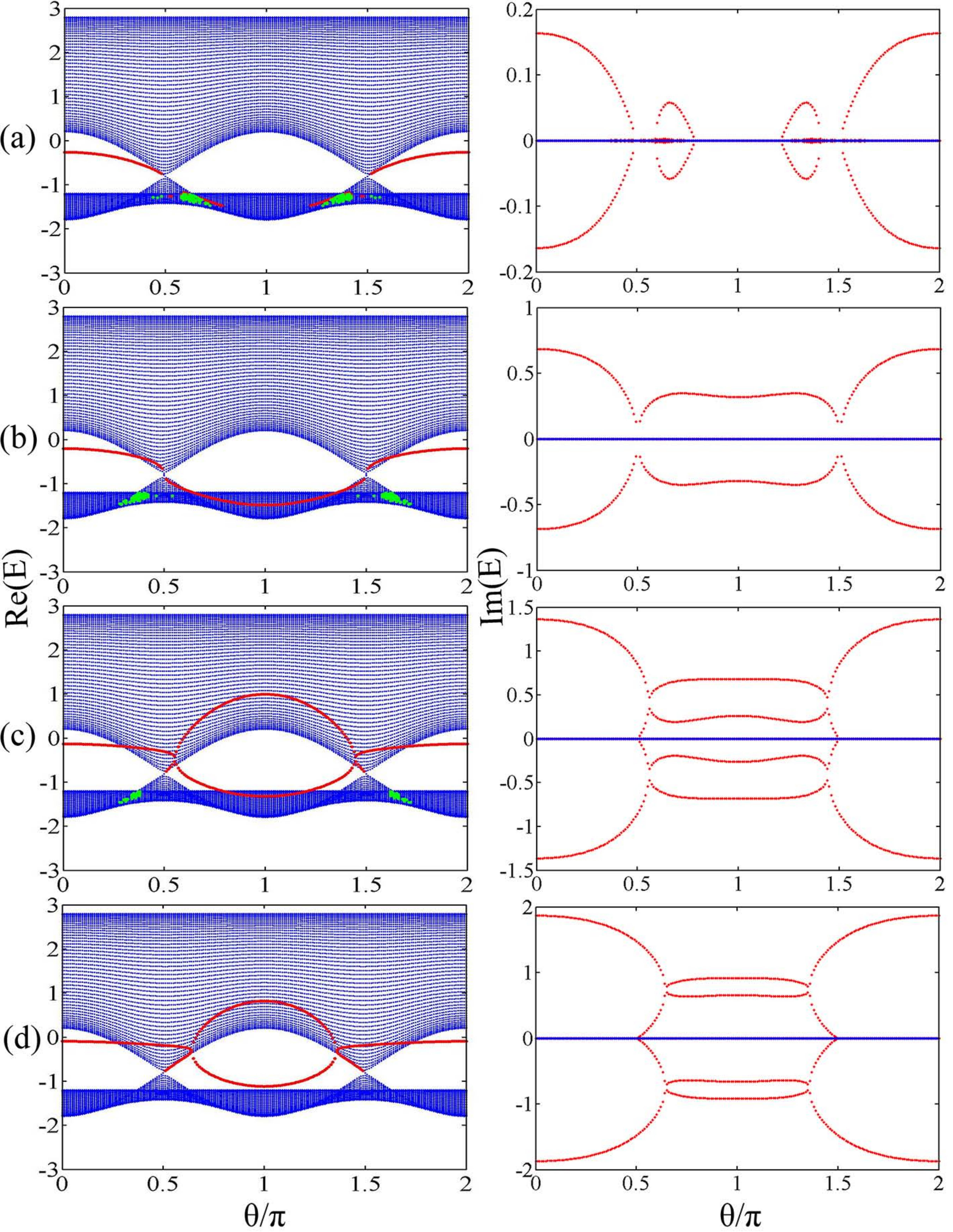}} \caption{
Real and imaginary parts of the eigenenergy spectra as a function of $\theta$ with parameters $\tau=0.4$ and $N=50$ for several values of $\gamma$.
(a) $\gamma=0.2$; (b) $\gamma=0.8$; (c) $\gamma=1.5$; (d) $\gamma=2.0$.}
\label{Fig6}
\end{figure}
Fig.6 shows the eigenenergy of the SSH model as a function of $\theta$, when the NNN intersite coupling is taken to be $\tau=0.4$. And in Fig.6(a)-(d), the strength of $\cal PT$-symmetric potentials is $\gamma=0.2$, $0.8$, $1.5$ and $2.0$ respectively. In addition to the band changes, we can clearly observe the variation of the topological states in the \emph{original} topologically trivial region, i.e., $\frac{1}{2}\pi<\theta<\frac{3}{2}\pi$. In the case of $\gamma=0.2$, the topological state begins to arise in the bonding band, along the two sides of this region. As a result, the type-I and type-IV $\cal PT$-symmetry breaking takes place in this region. When $\gamma=0.8$, one complete eigenenergy curve of the topological state can be found in the region of $\frac{1}{2}\pi<\theta<\frac{3}{2}\pi$. And the $\cal PT$-symmetry breaking mode is kept. On the other hand, with the further increase of imaginary potentials to $\gamma=1.5$, more complicated result comes into being. The topological state of the bonding band begins to rise and enter the interband gap, and also, the other complete topological state forms in the antibonding band. One then finds a loop-like real part of the toplogical-state energy to form in the middle area, corresponding to the type-II $\cal PT$-symmetry breaking. At the same time, two topological states near $\theta=\frac{1}{2}\pi$ and $\theta=\frac{3}{2}\pi$ appear, and they link the Dirac points and the EPs, respectively, which exactly means the type-III $\cal PT$-symmetry breaking. When $\gamma=2.0$, the region of type-II $\cal PT$-symmetry breaking decreases, while the type-III $\cal PT$-symmetry breaking becomes apparent. Such a complicated situation exactly describes the coexistence of the type-II, type-III and type-IV $\cal PT$-symmetry breaking, as well as the corresponding EP in the $\theta$ space.

\begin{figure}[htb]
\centering \scalebox{0.17}{\includegraphics{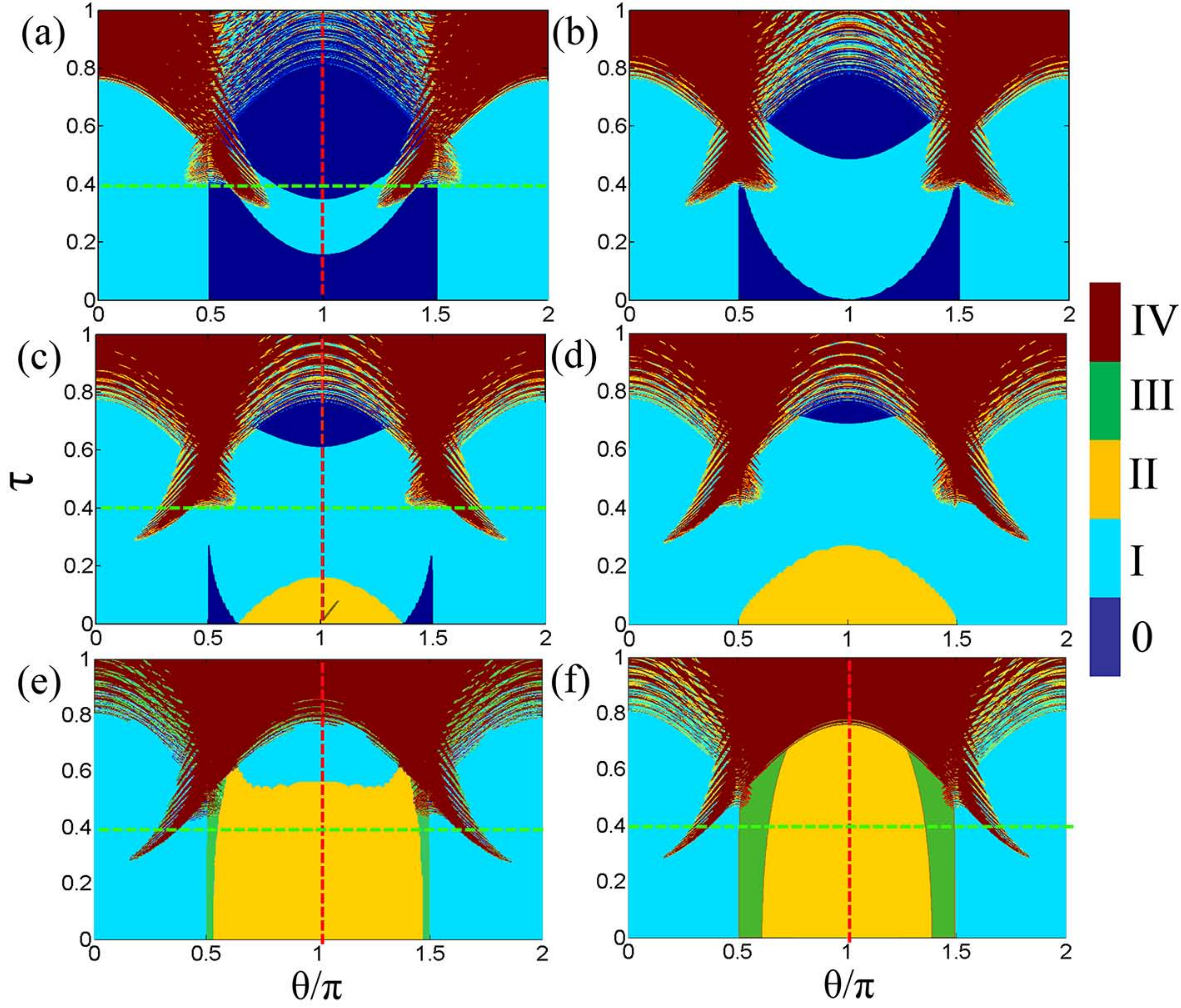}} \caption{
Phase diagram of the $\cal PT$-symmetry breaking caused by the increase of NNN intersite coupling and $\theta$. Relevant parameters are $\delta=0.5$ and $N=50$. (a) $\gamma=0.2$; (b) $\gamma=0.5$; (c) $\gamma=0.8$; (d) $\gamma=1.0$; (e) $\gamma=1.5$; (f) $\gamma=2.0$.}
\label{Fig7}
\end{figure}
The above results suggest that the topological properties of this system are accordant with the $\cal PT$-symmetry breaking. We then would like to present the more phase diagrams of the $\cal PT$-symmetry breaking to investigate the topological characteristics of it. The $(\theta,\tau)$ phase diagram is shown in Fig.~\ref{Fig7}, whereas the $(\tau,\gamma)$ result is in Fig.8. In Fig.7, the red and green dotted lines correspond to the parameters of Fig.5 and Fig.6, respectively. One can readily see in Fig.7(a)-(b) that in the cases of $\gamma\le0.5$, one topological state is allowed to arise in the region of $\frac{1}{2}\pi<\theta<\frac{3}{2}\pi$, with the occurrence of the type-I $\cal PT$-symmetry breaking. And the corresponding region is widened with the increase of $\gamma$. When the larger $\gamma$ is introduced, two topological states have opportunities to comes into being in the small-$\tau$ situation, which is manifested as the appearance of the II phase in the region of $\frac{1}{2}\pi<\theta<\frac{3}{2}\pi$. And such a phase is extended with the increase of $\gamma$, as shown in Fig.7(c)-(d). As a result, the 0-phase part is apparently suppressed. This exactly reflects the change of the $\cal PT$-symmetry breaking manner. Next, when the magnitude of $\cal PT$-symmetric potentials is further enhanced, the II-phase region is widened along the $\tau$-increase direction. And it is accompanied by the emergence of the III-phase near the boundary between the \emph{original} topologically trivial and nontrivial regions. With the further increase of the imaginary potentials, type-I phase in the \emph{original} topologically trivial region is absolutely suppressed, and the type-II and type-III phases occur in a larger scope of $\tau$ [see Fig.7(f)]. In addition to the $\cal PT$-symmetry breaking induced by the topological states, the bulk-state energies are also become complex, in the large-$\tau$ region. Therefore, the small $\tau$, e.g., $\tau<0.4$, is suitable to observe the effect of NNN intersite coupling on the topological properties of the SSH model as well as the EP and manner of $\cal PT$-symmetry breaking.

\par
\begin{figure}[htb]
\centering \scalebox{0.08}{\includegraphics{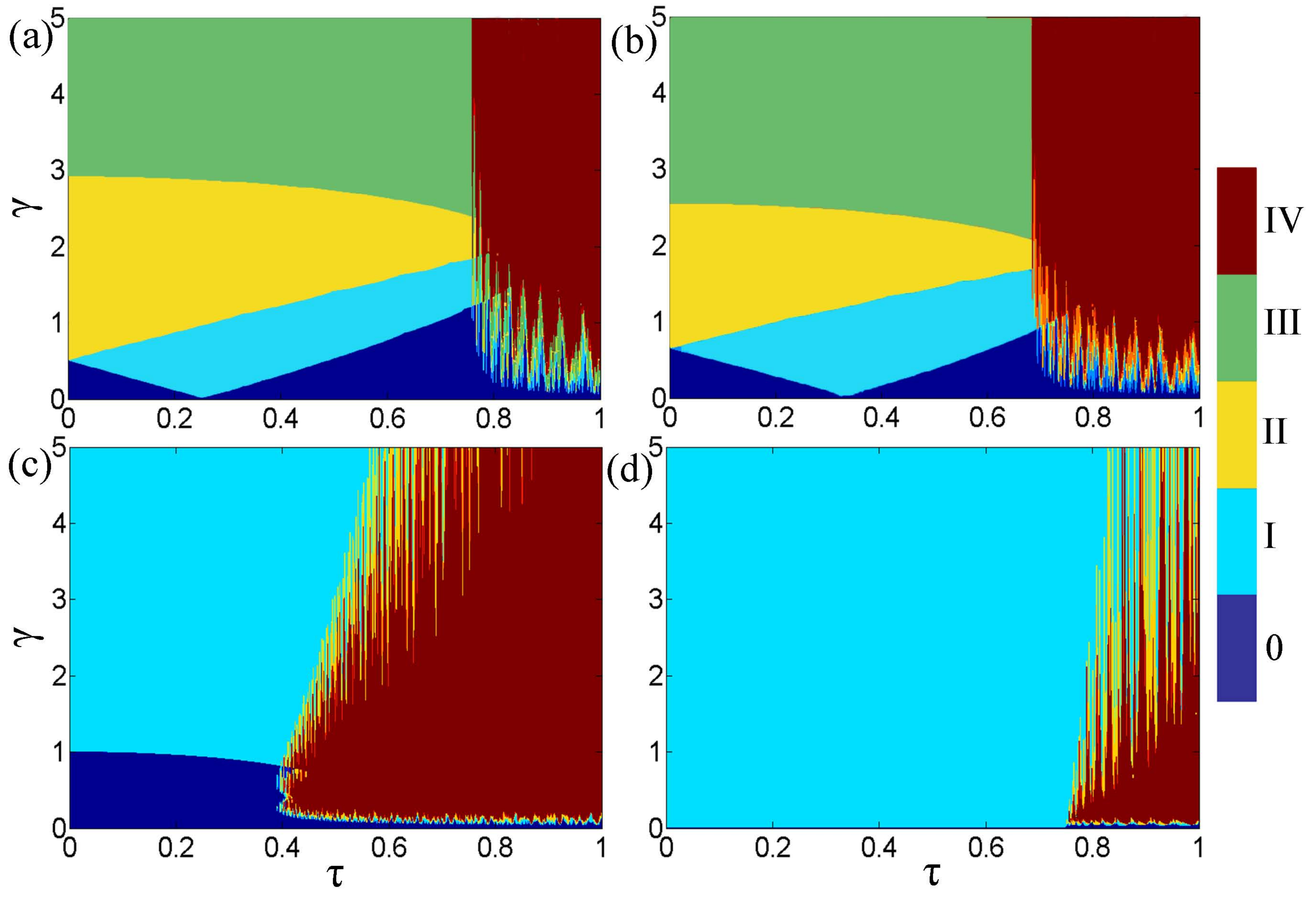}} \caption{
Phase diagram of the $\cal PT$-symmetry breaking of the SSH model, due to the increase of $\tau$ and $\gamma$. (a) $\theta=\pi$, (b) $\theta=1.25\pi$, (c) $\theta=1.5\pi$, (d) $\theta=2\pi$. The meanings of the colors are identical with those in Fig.~\ref{Fig7}.}
\label{Fig8}
\end{figure}
Fig.~\ref{Fig8} shows the phase diagram contributed by the increase of the NNN intersite coupling $\tau$ and the imaginary potential $\gamma$. As for $\theta$, we consider it to be $\pi$, $1.25\pi$, $1.5\pi$, and $\theta=2\pi$, respectively, in Fig.8(a)-(d). Thus Fig.~\ref{Fig8}(a)-(b) describe results in the topologically trivial region, Fig.~\ref{Fig8}(c) corresponds to the boundary between the topologically trivial and nontrivial regions, whereas Fig.~\ref{Fig8}(d) represents the topologically nontrivial region. One can then understand the roles of $\tau$ and $\gamma$ in modulating the topological properties and the $\cal PT$-symmetry breaking. According to Fig.~\ref{Fig8}(a)-(b), with the strengthening of the NNN intersite coupling, the 0 phase is first decreased and then increased gradually with the transition point at $\tau_c\approx 0.25$. And the I-phase region exhibits the opposite variation behavior, accompanied by its shift to the larger $\gamma$ direction. And it is clearly shown that the II-phase region is narrowed obviously, whereas the III-phase region exhibit the alternative result. The IV-phase occurs in the whole range of $\gamma$ with large $\tau$. Similar result can be observed in Fig.8(b). The difference lies in the right shift of the transition point $\tau_c$, the decrease of the II-phase region, and the increase of the regions of the other phases. Next in Fig.8(c), we find the type-I $\cal PT$-symmetry breaking when $\gamma$ increases to 1.0, in the case of $\tau\le0.4$. Once $\tau>0.4$, the IV phase becomes dominant. On the other hand, when $\theta=2\pi$, $\tau$ and $\gamma$ cannot change the type of $\cal PT$-symmetry breaking until $\tau\approx0.8$ where the type-IV $\cal PT$-symmetry breaking begins to arise, as shown in Fig.8(d). Up to now, we have known the roles of the NNN intersite coupling in modulating the characteristics of the $\cal PT$-symmetric non-Hermitian SSH structure.

\section{summary\label{summary}}
To summarize, we have investigated the eigenenergy characteristics of the $\cal PT$-symmetric non-Hermitian SSH model with two conjugated imaginary potentials at the end sites, by taking into account of the NNN intersite coupling. As a result, it has been found that such a coupling does not only enable to change the eigenenergies of both the bulk states, but also modifies the properties of the topological states in a substantial way. On the one hand, with the increase of NNN coupling, the bonding band is first narrowed and then widened with the top-bottom reversal, whereas the antibonding band is widened monotonously. On the other hand, the topological state in the non-Hermitian SSH model undergoes abundant properties. It shows that the real part of its energy departs from the energy zero point, and this state can also extend into the \emph{original} topologically-trivial region. Furthermore, new topological state emerges in such a region. These changes is accompanied by the new and complicated $\cal PT$-symmetry breaking, as described by the phase diagrams. All these phenomena should be attributed to the interplay between the particle-hole symmetry breaking and $\cal PT$-symmetric non-Hermitian nature. Therefore, the topological properties of the SSH model, as well as the $\cal PT$-symmetry breaking, can be efficiently modulated by considering the coexistence of the $\cal PT$-symmetry mechanism and NNN intersite coupling.
\section*{Acknowledgments}
W. J. G. thanks Hui Jiang and Qi Zhang for help discussions. This work was financially supported by the Fundamental Research Funds for the Central Universities (Grant Nos. N170506007 and N180503020), the Liaoning BaiQianWan Talents Program (Grant No. 201892126).

\clearpage

\bigskip

\end{document}